\documentclass[aps,prl,showpacs,twocolumn,amsmath,amssymb,
superscriptaddress,footinbib]{revtex4}

\usepackage{latexsym}
\usepackage{graphicx}
\usepackage{subfigure}
\usepackage{epsfig}
\usepackage{amsfonts}
\usepackage{amssymb}
\usepackage{amsmath}
\usepackage{bm}
\newcommand{\tr}{{\rm tr}\;}
\renewcommand{\i}{{\rm i}}

\usepackage{braket}
\begin{document}

\title{Exactly Solvable Fermion Chain Describing a $\nu=1/3$ Fractional
Quantum Hall State}

\author{Masaaki Nakamura}
\affiliation{Department of Physics, Tokyo Institute of Technology,
O-Okayama, Meguro-ku, Tokyo 152-8551, Japan}
\author{Zheng-Yuan Wang} 
\affiliation{Department of Physics, Tokyo Institute of Technology,
O-Okayama, Meguro-ku, Tokyo 152-8551, Japan}

\author{Emil J. Bergholtz}
\affiliation{Dahlem Center for Complex Quantum Systems
and Institut f\"ur Theoretische Physik,
Freie Universit\"at Berlin,
Arnimallee 14, 14195 Berlin, Germany}

\date{\today}

\begin{abstract}
We introduce an exactly solvable fermion chain that describes a
$\nu=1/3$ fractional quantum Hall (FQH) state beyond the thin-torus
limit.  The ground state of our model is shown to be unique for each
center of mass sector, and it has a matrix product representation that
enables us to exactly calculate order parameters, correlation functions,
and entanglement spectra. The ground state of our model shows striking
similarities with the BCS wave functions and quantum spin-1
chains. Using the variational method with matrix product ansatz, we
analytically calculate excitation gaps and vanishing of the
compressibility expected in the FQH state.  We also show that the above
results can be related to a $\nu=1/2$ bosonic FQH state.
\end{abstract}

\maketitle

{\it Introduction.---}
The fractional quantum Hall (FQH) effect is one of the fascinating
phenomena in condensed matter physics: In a 2D electron system in a
magnetic field, the Hall conductivity is quantized as
$\sigma_{H}=(e^2/h)\nu$ with the filling factor given by a rational
number $\nu=p/q$, due to strong electron-electron interactions
\cite{Tsui,Laughlin}.  Although three decades have past since its
discovery, the importance of this research field is still increasing,
partly due to new possible realizations of FQH phenomena including flat
band Chern insulators \cite{chernins} and bosonic systems of trapped
atoms \cite{Lin}.

In recent years, there have been theoretical efforts to study FQH states
in torus boundary conditions which can reduce the 2D continuum system in
a magnetic field to a 1D lattice model \cite{bk1,seidel}. This approach
sheds new light on the FQH physics, and is also used to analyze new type
of FQH states in flat band Chern insulators \cite{Qi}.

In this Letter, based on the 1D approach, we introduce a minimal model
with an exact ground state which describes a $\nu=1/3$ FQH state
[Eq.~(\ref{hamtrunc}) below].  We construct the Hamiltonian in terms of
local positive operators much like the Affleck, Kennedy, Lieb, and
Tasaki model for a quantum spin-1 chain \cite{Affleck-K-L-T}. We discuss
the properties of this model by obtaining exact expressions for various
correlation functions, order parameters, and entanglement spectra.
Moreover, excitation gaps are accurately obtained via variational
calculations.


{\it 1D description of FQH states.---} 
We consider 2D interacting electrons in a magnetic field $B$ on toroidal
boundary conditions, where $L_i$ ($i=1,2$) are the circumferences of the
torus of the corresponding coordinates $x_i$, and $l_{B}\equiv
\sqrt{\hbar/eB}$ is the magnetic length which will be set to unity.  As
discussed in preceeding works \cite{bk1,seidel}, the system with
two-body interaction assumes the following 1D discretized model,
\begin{equation}
\mathcal{H}=
\sum_{i=1}^{N_s}
\sum_{k>|m|}
V_{km}
 c_{i+m}^{\dag}
 c_{i+k}^{\dag}
 c_{i+m+k}^{\mathstrut}
 c_{i}^{\mathstrut},
 \label{1D_model}
\end{equation}
where $c_i^\dagger$ ($c_i$) creates (destroys) a fermion at site $i$,
and the number of lattice sites is given by $N_s=L_1 L_2/2\pi$.  The
matrix-element $V_{km}$ specifies the amplitude for a process where two
particles with separation $k+m$ hop $m$ steps to opposite
directions. This process conserves the center-of-mass coordinate
$K_1\equiv \sum_{i=1}^{N_s} i \hat{n}_i$ (mod $N_s$) with $\hat
n_i\equiv c_i^\dagger c_i^{\mathstrut}$, which corresponds to the
momentum along $x_1$ direction.  Therefore, the system with $\nu=p/q$
can be divided into $q$ independent subsystems.

For the pseudo potential \cite{Trugman-K} which has the $\nu=1/3$
Laughlin wave function \cite{Laughlin} as an exact ground state, the
matrix elements for large $L_2$ are
\begin{equation}
 V_{km}\propto(k^2-m^2)e^{-2\pi^2(k^2+m^2)/L_1^2}.
  \label{m_elements}
\end{equation}
Thus the hopping terms ($m\neq 0$) are suppressed exponentially compared
to the electrostatic terms ($m=0$) in the thin-torus (TT) limit $L_1\to
0$, and the system becomes a charge-density-wave state
$\ket{\Psi_0}=\ket{100100100\cdots}$.  In order to describe systems with
finite $L_1$, we include also the leading hopping terms.  This expansion
in $e^{-2\pi^2/L_1^2}$ is well controlled, and we expect it to capture
the physics also for more general interactions
\cite{bk1,bk23,seidel,rh94,Wang-S}.


{\it Model with exact ground state.---} 
Based on the above framework, we truncate the long-range interactions of
the 1D model (\ref{1D_model}) at $\nu=1/3$ up to the third neighbor
($k+m\leq 3$) assuming only $\sqrt{V_{10}V_{30}}=V_{21}$ which is
satisfied in Eq.~(\ref{m_elements}), $\forall L_1$. Then we have
\begin{align}
 \mathcal{H}=
 &\sum_{i=1}^{N_s} [
 \alpha_i^2\hat n_{i+1} \hat n_{i+2}
 +\beta_i^2\hat n_i \hat n_{i+2}
 +\gamma_i^2\hat n_{i} \hat n_{i+3}\nonumber \\ 
 &+\alpha_i\gamma_i
 (c^\dagger_{i} c^\dagger_{i+3}
 c_{i+2}^{\mathstrut} c_{i+1}^{\mathstrut} + {\rm H.c.})],
 \label{hamtrunc} 
\end{align}
where we have generalized the parameters $\alpha_i,\beta_i,\gamma_i\in
{\mathbf R}$ to have site dependence.  This truncation is valid as an
expansion beyond the TT limit \cite{Note1}. Now we rewrite this model in
the following form
\begin{equation}
 \mathcal{H}=
  \sum_{i} [Q^\dagger_i Q_i^{\mathstrut}
  + P^{\dagger}_i P_i^{\mathstrut}]
  \ ,
  \label{hamproj}
\end{equation}
where
\begin{equation}
 Q_i=\alpha_i c_{i+1}c_{i+2}+\gamma_i c_{i}c_{i+3},\quad
  P_i=\beta_i c_{i}c_{i+2}.
\end{equation}
Eq.~(\ref{hamproj}) is clearly a sum of positive operators, thus, the
spectrum is positive semidefinite $\braket{\mathcal{H}}\geq 0$. As we
will show, the (unnormalized) ansatz,
\begin{eqnarray}
 \ket{\Psi_{1/3}}=\prod_i(1 - t_i \hat{U}_i)\ket{\Psi_0}
  = \prod_i e^{-t_i\hat{U}_i}\ket{\Psi_0},
  \label{psi13} 
\end{eqnarray}
where $t_i\equiv\gamma_i/\alpha_i$ and $\hat{U}_i\equiv c^\dagger_{i+1}
c^\dagger_{i+2} c_{i+3}^{\mathstrut} c_{i}^{\mathstrut}$, provides the
unique zero-energy solutions. Note that $[\hat{U}_i,\hat{U}_j]=0$ for
$|i-j|\neq 2$ and $\hat{U}_i\hat{U}_{i+2}\ket{\Psi_0}=0$.  In
Eq.~(\ref{psi13}), the original state $\ket{\cdots 1001\cdots}$ in
$\ket{\Psi_0}$ and its ``squeezed'' counterpart $\ket{\cdots
0110\cdots}$ cancel each other by acting $Q_i$, and there are no
next-nearest pairs $\ket{\cdots 101\cdots}$.  Hence this state satisfies
$Q_i\ket{\Psi_{1/3}}=P_i\ket{\Psi_{1/3}}=0, \forall i$, and it is a zero
energy ground state \cite{Jansen}.  Due to the conservation of the
center of mass, this state has threefold degeneracy for periodic
systems, even when the parameters have site dependence.
Our wave function (\ref{psi13}) gives exact ground states for open
boundary systems.  We can obtain many new zero-energy eigenstates at a
lower filling than $\nu=1/3$ by inserting an extra $0$ anywhere in the
root state, $\ket{\Psi_0}$ in (\ref{psi13}), because the insertion of
$0$ is equivalent to make open boundaries.  Moreover, if $\cdots 000101$
type configurations are located at these ``edges'', they also give
eigenstates with finite energies due to the $\beta_i^2$ term.

The uniqueness of the ground state for each center-of-mass sector in the
$\nu=1/3$ periodic case can be shown using the Perron-Frobenius theorem:
First, one can show that all the states generated by acting with the
Hamiltonian on $\ket{\Psi_0}$ can be reached by successive applications
of $\hat{U}_i$ (for different $i$) \cite{Wang-T-N}; thus, the
Hamiltonian takes the form of a connected matrix in this subspace. Next,
with a unitary transformation that changes the signs of $\gamma_i$, all
the off-diagonal matrix elements can be made negative. Now the
Perron-Frobenius theorem implies that there are no other zero-energy
states than (\ref{psi13}) in this subspace, since all its expansion
coefficients have the same sign.  Finally, it follows that all states
that are not connected to $\ket{\Psi_0}$ (or translations thereof) by
the Hamiltonian always include finite amplitudes of next-nearest
neighbor particle which costs energy when $\beta_i^2>0$; thus, all such
states have a finite energy.  This concludes the proof that
(\ref{psi13}) is the unique ground state up to translations.


{\it Correlation functions and order parameters.---}
From the exact solution (\ref{psi13}) it is possible to calculate rather
generic quantities such as correlation functions and entanglement
properties of the ground state.  For this purpose, it is convenient to
introduce a matrix product (MP) representation
\cite{Fannes-N-W,Klumper-S-Z} of the ground state wave function
(\ref{psi13}).  In a periodic system with $N_s=3N$ sites, the normalized
ground state wave function (\ref{psi13}) can be written as
\begin{equation}
 \ket{\Psi_{1/3}}={\cal N}^{-1/2}\tr[g_1 g_2 \cdots g_N],
  \label{MPS}
\end{equation}
where $g_j$ is identified as the following $2\times 2$ matrix,
\begin{equation}
 g_j=\left[
      \begin{array}{c c}
       |{\rm o} \rangle_j &  |+ \rangle_j \\
       -t_{3j}|- \rangle_j  & 0
      \end{array}
     \right].
 \label{g-reduced}
\end{equation}
Here we have introduced the spin-1 representation for three-sites unit
cell \cite{Nakamura-B-S}: $\ket{010}\to\ket{\rm o}$, $\ket{001}\to\ket{+}$,
and $\ket{100}\to\ket{-}$.  For the open boundary system, we should
extract only $(1,1)$ component of (\ref{MPS}) instead of taking the
trace.

\begin{figure}[t]
\centerline{\includegraphics[width=0.80\linewidth]{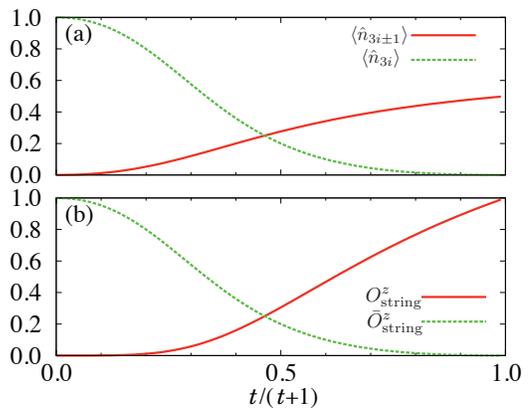}}
 \caption{(a) Density functions $\braket{\hat{n}_i}$ in three-sites unit
 cell, and (b) String order parameter $O_{\rm string}^z$ and dual string
 order parameter $\bar{O}_{\rm string}^z$ as functions of $t$.  $O_{\rm
 string}^z$ ($\bar{O}_{\rm string}^z$) dominant in the large (small) $t$
 regime plays a role to characterize the ``superconducting''
 (``normal'') component in an analogy of the BCS
 theory.}\label{fig:string}
\end{figure}

Using the MP formalism for uniform $t_i=t$ and infinite systems
$N\to\infty$, we obtain the density function which has three-site
periodicity as [see Fig.~\ref{fig:string}(a)],
\begin{equation}
 \braket{\hat{n}_{3i\pm1}}=
\frac 1 2 \left(1-\frac 1{\sqrt{4t^2+1}}\right)
,\,
  \braket{\hat{n}_{3i}}=
  \frac 1{\sqrt{4t^2+1}}\ .
  \label{dens_func}
\end{equation}
This result shows that the density function becomes uniform at
$t=\pm\sqrt{2}$ \cite{Note2}. The single particle correlation function
is given by
$\braket{c^{\dag}_{i}c_j^{\mathstrut}}=\delta_{ij}\braket{\hat{n}_i}$
due to the center-of-mass conservation.

We can also obtain density-density correlation functions in a similar
way. In the infinite-size limit we find exponentially decaying
correlations
\begin{equation}
 \braket{n_{i}n_{j}}\!-\!\braket{n_{i}}\braket{n_{j}}\!\sim\! 
  \left(\!\frac{1-\sqrt{4t^2+1}}{1+\sqrt{4t^2+1}\!}
  \right)^{|i-j|/3}\!\equiv e^{-|i-j|/\xi}.
  \label{corr_func}
\end{equation}
As $|t|\rightarrow\infty$, the correlation length $\xi$ diverges which
reflects the state $\ket{+-+-\cdots}+\ket{-+-+\cdots}$.

Using the suggestive analogy of quantum spin-1 chains, we consider
nonlocal order in terms of the string order parameters $O_{\rm
string}^\alpha=-\braket{S^\alpha_ie^{\i\pi\sum_{k=i+1}^{j-1}
S^\alpha_k}S^\alpha_j}$ \cite{Nijs-R}. This is known to characterize the
Haldane-gap (including N\'eel) state.  On the other hand, a ``dual''
string order parameter $\bar{O}_{\rm
string}^\alpha=\braket{e^{\i\pi\sum_{k=i+1}^{j-1} S^\alpha_k}}$
\cite{Berg-T-G-A} can be introduced to characterize the large-$D$
phases.  Using the MP formalism we find
\begin{equation}
 O_{\rm string}^z
  =\frac{(\sqrt{4t^2+1}-1)^2}{4t^2+1},\quad
  \bar{O}_{\rm string}^z
  =\frac{1}{4t^2+1},
  \label{SOPs}
\end{equation}
where $N$, $j-i\to\infty$ is assumed, and the $x,y$ components are
vanishing. In conventional quantum spin chains, these two order
parameters are usually not finite simultaneously. However, the present
spin-mapped system breaks the space-inversion and spin-reversal
symmetries (e.g., a configuration $\ket{\cdots + -\ {\rm o}\cdots}$
occur in (\ref{MPS}) but $\ket{\cdots {\rm o} - +\cdots}$ does not), and
also SU(2) symmetry, which enables the two different orders to coexist.
This results consistent with the numerical analysis which concludes that
the spin mapped $\nu=1/3$ FQH state is adiabatically connected both from
``Haldane''(N\'eel) and large-$D$ phases without closing the energy gap
\cite{Nakamura-B-S, Berg-T-G-A,Pollmann-T-B-O}.  The two string order
parameters behave as $O_{\rm string}^z\to 1$ for $|t|\to\infty$ and
$\bar{O}_{\rm string}^z\to 1$ for $t\to 0$ as shown in
Fig.~\ref{fig:string}(b).  This can be interpreted as the two limits
characterize ``superconducting'' and ``normal'' states in analogy with
the BCS wave function which is very similar to Eq.~(\ref{psi13})---it
has the form of bosonic operators acting on a vacuum state.

\begin{figure}[t]
\includegraphics[width=0.82\linewidth]{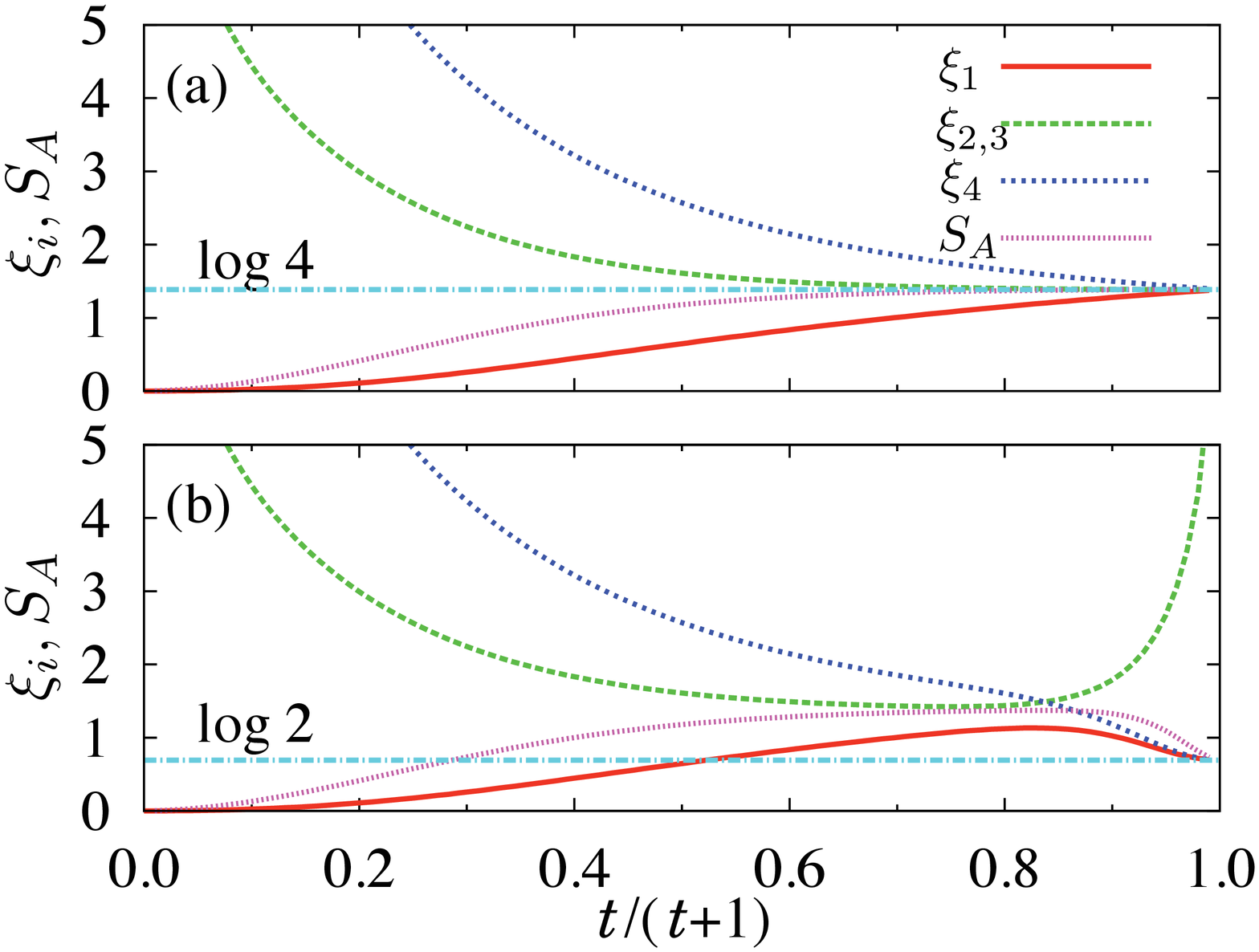}
 \caption{Entanglement spectra $\{\xi_i\}$, and entanglement entropy
 $S_A$, as functions of $t$ for (a) infinite-size system and (b)
 finite-size ($N=2L=32$) system.}\label{fig:entangle}
\end{figure}


{\it Entanglement spectrum and entropy.---} 
We can also derive the entanglement spectrum (ES) $\{\xi_i\}$
\cite{Li-H} of the system in the spin-1 MP basis via the Schmidt
decomposition dividing the system into two parts, $\{k_1,\ldots,k_L\}\in
A$ and $\{k_{L+1},\ldots,k_N\}\in B$, as
\begin{align}
 \ket{\Psi_{1/3}}
 &={\cal N}^{-1/2}
 \sum_{j_1, j_2=1,2} 
 \left\{
 \left[ g_{1}\cdots g_{L} \right]_{j_1 j_2}
 \left[ g_{{L+1}}\cdots g_{N} \right]_{j_2 j_1} \right\}\nonumber\\&
 =\sum_{j_1, j_2=1,2} 
 (\mathcal{N}^A_{j_1,j_2}\mathcal{N}^B_{j_1,j_2}/\mathcal{N})^{1/2}
 \ket{\psi^A_{j_1,j_2}}\otimes\ket{\psi^B_{j_1,j_2}}
 \nonumber\\&
 \equiv\sum_ie^{-\xi_i/2}|\psi_{i}^A\rangle\otimes|\psi_{i}^B\rangle
\end{align}
where $|\psi_{i}^{A(B)}\rangle$ with $i\equiv (j_1,j_2)$ are orthogonal
states describing subsystem $A(B)$, and $\mathcal{N}^{A(B)}_{j_1,j_2}$
are their norms.  In the infinite-size limit, $L=N/2 \rightarrow
\infty$, one finds $\xi_{1},\xi_{4}=\log(4+t^{-2})
\pm\log\left(\frac{\sqrt{4t^2+1}-1}{\sqrt{4t^2+1}+1\!}\right)$ and
$\xi_{2}=\xi_{3}=\log(4+t^{-2})$ [see Fig.~\ref{fig:entangle}(a)].
The structure of the ES is different from that of usual Haldane-gap
systems characterized by two-fold degeneracy in all ES
($\xi_{1}=\xi_{4}$ and $\xi_{2}=\xi_{3}$) \cite{Pollmann-B-T-O}. This is
because our ``Haldane'' state is rather close to a N\'eel state which
does not have edge spins, due to the lacking of the symmetries.
We also get the von Neumann entanglement entropy $S_A
=\sum_{i}\xi_{i}e^{-\xi_{i}}$ which approaches $\log 4$ for large
$t$. The finite entanglement entropy is a generic property of 1D gapped
states \cite{arealaw}.

In finite systems the entanglement properties are somewhat altered as
shown in Fig.~\ref{fig:entangle}(b). Once the correlation length $\xi$
is of the order of the distance between the cuts $L$, the above
structure of the ES breaks down \cite{Lauchli} and for large enough $t$
the entanglement entropy is instead approaching $\log 2$ due to the
states $\ket{+-+-\cdots}$ and $\ket{-+-+\cdots}$.


\begin{figure}[t]
 \includegraphics[width=0.88\linewidth]{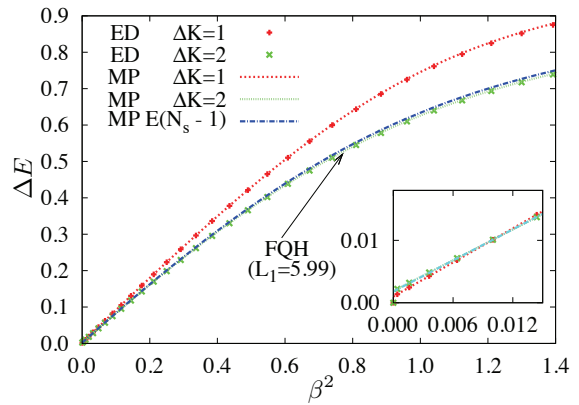}
 \caption{The lowest excitation gaps of the model (\ref{hamtrunc}) at
 $\alpha=1$, $t=1/3$. The neutral gaps for $\Delta K=1,2$ are obtained
 by exact diagonalization (ED) for $N_s=27$ (dots) and the variational
 calculation with the MP ansatz for $N_s\to\infty$ (lines).  $E(N_s-1)$
 for $N_s\to\infty$ is also obtained by the MP ansatz.  Here,
 $\beta^2=0.77$ corresponds to a FQH system on a torus with
 circumference $L_1=5.99$.}  \label{fig:excitation}
\end{figure}

{\it Compressibility and excitation gaps.---}
When we shrink the system size as $N_s\to N_s-1$ by removing $0$ from
the root state, a $10$-type domain wall appears that carries a
fractional charge \cite{Laughlin} $e^*=e/3$ (The fractional charge
follows from noting that creating three such domain walls $101010$
amounts to adding one electron to the root state $100100$).
This excitation energy $E(N_s-1)$ can be analytically calculated within
a variational approach based on the MP formalism.  Considering a
subspace given by (\ref{psi13}) where $\ket{\Psi_0}$ is replaced by
$\ket{\Psi_0^-}=\ket{10010100100\cdots}$, and appropriate local
deformation of the MP state, we get a finite value of $E(N_s-1)$ for
$N_s\to\infty$ as shown in Fig.~\ref{fig:excitation}. Since
$E(N_s+1)=E(N_s)=0$ as already discussed, we obtain divergence of the
inverse compressibility as expected for the FQH state,
\begin{align}
 \kappa^{-1}=\lim_{N_s\to\infty}N_s
 \frac{E(N_s-1)+E(N_s+1)-2E_s(N_s)}{4\pi l_{B}^2}\to\infty.
\end{align}

The excitation energies in the charge neutral sector can also be
calculated similarly. We specify these states by $\Delta K$ which means
the center-of-mass coordinate relative to the ground state.  Considering
$\Delta K=1$ and $\Delta K=2$ subspaces given by (\ref{psi13}) where
$\ket{\Psi_0}$ is replaced by states
$\ket{\Psi_1}=\ket{100100010100\cdots}$ and
$\ket{\Psi_2}=\ket{100010010100\cdots}$, we get excellent agreement with
the numerical results of the exact diagonalization which are very
insensitive to the system size, as shown in Fig.~\ref{fig:excitation}.
For $t=1/3$, the $\Delta K=1$ state is the lowest excitation only in the
very small $\beta$ region, while the $\Delta K=2$ state becomes the
lowest as $\beta$ is increased.  The tiny deviation at $\beta=0$ is due
to phase separated states, $\ket{1010\cdots 0000}$, which have zero
energy in this limit. The lowest $\Delta K=0$ excitation is
significantly higher in energy. Since the above features are
qualitatively unchanged from small to sufficiently large $t$ regions
($|t|\sim 1$), we identify the $\Delta K=2$ excitation gap as the
neutral
energy gap of a $\nu=1/3$ FQH state with toroidal boundary
conditions. This result is consistent with a recent analysis using the
spherical geometry and the Jack polynomials which identifies the neutral
gap in the $L=2$ angular momentum sector \cite{Yang-H-P-H}.


{\it Bosonic systems.---}
The present exact argument can also be applied to bosonic systems. The
Hamiltonian (\ref{hamproj}) with the operators $Q_i=\alpha_i
b_{i}b_{i}+\gamma_i b_{i-1}b_{i+1}$ and $P_i=\beta_i b_{i}b_{i+1}$
defines a $\nu=1/2$ bosonic FQH state that has the following exact
two-fold degenerate ground state:
\begin{equation}
 \label{psi12b}
  \ket{\Psi_{1/2}^{\rm B}}=\prod_i
  (1 - {\textstyle \frac{t_{i}}{\sqrt{2}}}b^\dagger_{i+1}b^\dagger_{i+1}
  b_{i+2}^{\mathstrut} b_{i}^{\mathstrut})\ket{\Psi_0^{\rm B}},
\end{equation}
where $\ket{\Psi_0^{\rm B}}\equiv\ket{0101010\cdots}$ and
$t_i\equiv\gamma_i/\alpha_i$.  In this model the spin-1 mapping is also
possible as $\ket{10}\to \ket{\rm o}$, $\ket{02}\to \ket{+}$, and
$\ket{00}\to \ket{-}$. Hence, we obtain the same effective spin-1
representation (\ref{MPS}) as in the $\nu=1/3$ fermion system.


{\it Conclusion.---}
We have introduced a 1D interacting fermion model with an exact ground
state that describes a $\nu=1/3$ FQH state.  We have demonstrated the
uniqueness of the ground state with periodic boundary conditions for
each center-of-mass sector.  We have introduced a MP representation of
the ground states and obtained exact expressions for various correlation
functions, order parameters and the entanglement spectra. Moreover, the
excitation gaps have been accurately obtained via variational
calculations with MP ansatz.  We have also shown that the present
argument can be applied to a $\nu=1/2$ bosonic FQH state.  Although our
thin-torus approach does not describe a genuine liquid state
\cite{rh94}, it captures other important aspects of the FQH physics.


{\it Acknowledgments.---} We thank Juha Suorsa for useful discussions
and for related collaborations.  M.~N.  and E.~J.~B.  acknowledge the
visitors program of the Max Planck Institute f\"{u}r Physik komplexer
Systeme, Dresden, where this work was initiated.  M.~N. acknowledges
support from Global Center of Excellence Program ``Nanoscience and
Quantum Physics'' of the Tokyo Institute of Technology and Grants-in-Aid
No.23540362 by MEXT.  E.~J.~B. was supported by the Alexander von
Humboldt Foundation.


\end{document}